\documentclass[sigconf,natbib=false, nonacm]{acmart}
\AtBeginDocument{%
  }

\setcopyright{acmlicensed}
\copyrightyear{}
\acmYear{2026}
\acmDOI{XXXXXXX.XXXXXXX}
\acmConference[MSR '26]{23rd International Mining Software Repositories Conference}{April 13--14,
  2026}{Rio de Janeiro, Brazil}
\acmISBN{978-1-4503-XXXX-X/2018/06}

\definecolor{blue}{HTML}{000000}

\RequirePackage[
  datamodel=acmdatamodel,
  style=acmnumeric,
  ]{biblatex}

\usepackage{algorithm}
\usepackage{algorithmicx}
\usepackage{algcompatible}
\usepackage{alltt}

\begin{document}

\title{Beyond the Prompt: Assessing Domain Knowledge Strategies for High-Dimensional LLM Optimization in Software Engineering}

\author{Srinath Srinivasan}
\email{ssrini27@ncsu.edu}
\affiliation{%
  \institution{North Carolina State University}
  \city{Raleigh}
  \state{North Carolina}
  \country{USA}
}

\author{Tim Menzies}
\email{timm@ieee.org}
\affiliation{%
  \institution{North Carolina State University}
  \city{Raleigh}
  \state{North Carolina}
  \country{USA}
}

\renewcommand{\shortauthors}{Srinivasan et al.}

\begin{abstract}
\textbf{Background/Context}: Large Language Models (LLMs) demonstrate strong performance on low-dimensional software engineering optimization tasks ($\le$11 features) but consistently underperform on high-dimensional problems where Bayesian methods dominate. A fundamental gap exists in understanding how systematic integration of domain knowledge
(whether from humans or automated reasoning) can bridge this divide.

\textbf{Objective/Aim}: We compare human versus artificial intelligence strategies for generating domain knowledge. We systematically evaluate four distinct architectures to determine if structured knowledge integration enables LLMs to generate effective warm starts for high-dimensional optimization.

\textbf{Method}: We evaluate four approaches on MOOT\footnote{Available at: \href{https://tiny.cc/moot}{tiny.cc/moot}} datasets stratified by dimensionality: (1) \textit{Human-in-the-Loop Domain Knowledge Prompting (H-DKP)}, utilizing asynchronous expert feedback loops; (2) \textit{Adaptive Multi-Stage Prompting (AMP)}, implementing sequential constraint identification and validation; (3) \textit{Dimension-Aware Progressive Refinement (DAPR)}, conducting optimization in progressively expanding feature subspaces; and (4) \textit{Hybrid Knowledge-Model Approach (HKMA)}, synthesizing statistical scouting (TPE) with RAG-enhanced prompting. Performance is quantified via Chebyshev distance to optimal solutions and ranked using Scott-Knott clustering against an established baseline for LLM generated warm starts.

Note that all human studies conducted as part of this study will comply with the policies of our local Institutional Review Board.
\end{abstract}



\keywords{Optimization, warm starts, LLM, active learning, configuration}


\maketitle

\section{Introduction}

Software engineering optimization requires balancing competing objectives (e.g. runtime versus memory, quality versus cost, speed versus energy consumption). Active learning addresses these problems efficiently by using models to select the most informative examples to label, achieving good results with minimal data \cite{senthilkumar2024largelanguagemodelsimprove}. However, to learn good optimizations from data, some ground truth data must initially be available. In the domain of software engineering, obtaining these labels is often prohibitively expensive or time-consuming~\cite{DBLP:conf/wosp/ValovPGFC17, krishna2020whence}.

Large Language Models (LLMs) offer a potential solution for generating these initial ``warm start'' samples without the cost of execution. Yet, their reliability in this domain is debated; Treude et al.~\cite{ahmed2025llmsreplacemanualannotation} conclude that LLMs function best as assistive rather than authoritative agents in software tasks. Conversely, insights from Nair et al.~\cite{nair2017using} on ``bad learners'' suggest that even imperfect models can successfully guide optimization if they can suggest valid partial rankings. This raises a critical question: can we leverage the assistive nature of LLMs to act as heuristic guides for optimization, even if they lack authoritative ground truth?

Recent work \cite{senthilkumar2024largelanguagemodelsimprove} studying dozens of SE optimization problems demonstrates that LLMs can indeed generate effective warm starts, reducing labeling requirements from hundreds to dozens of examples. On low-dimensional problems ($<6$ features), LLM-based warm starts achieve top performance 100\% of the time. On medium-dimensional problems (6-11 features), they succeed 50\% of the time. However, performance collapses for high-dimensional problems ($>11$ features), where traditional Bayesian methods like Gaussian Process Models remain superior. This dimensional barrier is problematic: many real-world SE optimization tasks in the MOOT \cite{mootrepo} repository involve more than 11 features, including software configuration (18-38 features), process modeling (23 features), and hyperparameter tuning (14-38 features).

The cause of this failure likely stems from training data limitations. LLMs excel on ``common'' problems well-represented in their training corpus but struggle with specialized, high-dimensional SE tasks that lack abundant public examples. Cloud configuration, aerospace software processes, and domain-specific optimizations rarely have publicly documented optimal solutions. Without sufficient training data, LLMs cannot learn the complex feature interactions that characterize high-dimensional spaces. This problem will intensify: recent projections suggest LLMs will exhaust available textual training data by 2028, making data-scarce optimization domains increasingly important~\cite{villalobos2024rundatalimitsllm}. No prior work has systematically explored whether injecting domain-specific knowledge can overcome these limitations.

We propose four complementary approaches to incorporate domain knowledge into LLM warm starts, moving from human-centric to fully automated strategies:
\begin{enumerate}
    \item \textbf{Human-in-the-Loop Domain Knowledge Prompting (H-DKP):} We leverage human experts to verify and refine constraints in an iterative loop, augmenting prompts with structured feature relationships and heuristics at three levels of detail.
    \item \textbf{Adaptive Multi-Stage Prompting (AMP):} We decompose the generation process into sequential reasoning stages: analysis, constraint identification, generation, and validation. We allow the LLM to generate its own knowledge before proposing solutions.
    \item \textbf{Dimension-Aware Progressive Refinement (DAPR):} We address high-dimensional complexity by optimizing in reduced feature spaces (identified via statistical ranking) and progressively expanding dimensions, effectively guiding the LLM through the search space.
    \item \textbf{Hybrid Knowledge-Model Approach (HKMA):} We employ Retrieval-Augmented Generation (RAG) \cite{lewis2020retrieval} combined with rapid statistical exploration using Tree of Parzen Estimators (TPE) \cite{NIPS2011_86e8f7ab} to leverage both data-driven patterns and semantic understanding from online texts and documentation.
\end{enumerate}

We will evaluate all approaches on MOOT multi-objective SE optimization tasks stratified by dimensionality. 

\section{Related Work}

Modern software engineering is characterized by massive parameter search spaces that must be configured~\cite{chen2018sampling}. Xu et al.~\cite{Xu15a} and Van Aken et al.~\cite{VanAken2017} report that as systems mature, these spaces explode exponentially. Consequently, users often ignore configuration options or rely on obsolete defaults, which can lead to significant performance degradations (up to 480$\times$ in some industrial cases~\cite{Zhou2011,Herodotou11,Jamshidi2016proposal}). To manage this complexity, researchers employ configuration optimization algorithms to balance competing constraints~\cite{harman2012search}. In the realm of software analytics, such optimization is critical; studies by Fu et al.~\cite{fu2016tuning} and Agrawal et al.~\cite{agrawal2019dodge} demonstrate that optimizing learners (e.g., finding hyperparameter settings) can fundamentally alter experimental conclusions, turning a "worst" performing algorithm into the "best"~\cite{Tantithamthavorn16, agrawal2018better, yedida2021value, yedida2023find}.

Despite the necessity of optimization, practical application is hindered by the high cost of data collection. Exploring configuration landscapes for systems like x264 can require thousands of hours of compile time~\cite{DBLP:conf/wosp/ValovPGFC17}, necessitating low-resource approaches that operate within strict budgets of fewer than 50 evaluations~\cite{DBLP:journals/tse/Nair0MSA20,DBLP:conf/icse/0003XC021}. To maximize efficiency within these limits, researchers utilize iterative refinement algorithms~\cite{mkaouer2015many,zhang2017constraint} and Active Learning \cite{rayegan2025minimal}. However, these methods are sensitive to initialization; poor "cold starts" waste the limited labeling budget. The standard solution has historically been "warm starting" the optimization using prior knowledge from Subject Matter Experts (SMEs)~\cite{hacohen2022active, liu2024large, yehuda2022active}.

While effective, reliance on human SMEs is not scalable across the diverse ecosystem of modern software libraries. Recent work attempted to use Large Language Models (LLMs) as automated proxies for SMEs to generate these warm starts \cite{senthilkumar2024largelanguagemodelsimprove}. However, these initial studies reveal a critical "dimensional barrier": while LLMs perform well on simple, low-dimensional tasks, their performance collapses on high-dimensional, multi-objective tabular data ($>11$ features), often performing worse than random sampling.

\section{Research Questions}

\begin{itemize}
    \item \textbf{RQ1 (Comparative Efficacy):} Which domain knowledge integration strategy (HDKP, AMP, DAPR, HKMA) yields the highest quality warm starts compared to standard baselines across the MOOT corpus?
    
    \item \textbf{RQ2 (The Dimensional Barrier):} How does the effectiveness of each strategy vary across dimensionality tiers?
    
    \item \textbf{RQ3 (The Human Factor):} Does human-in-the-loop feedback (H-DKP) provide statistically significant performance gains over the fully automated methods?
    
    \item \textbf{RQ4 (Cost-Benefit Analysis):} What are the quantitative trade-offs between solution quality improvements and computational overhead for each approach?
    
    \item \textbf{RQ5 (Knowledge Attribution):} Which specific categories of domain knowledge (structural constraints, feature correlations, heuristics, or statistical priors) contribute most to performance improvements?
\end{itemize}

\section{Data Sources \& Collection}

\subsection{The MOOT Repository}
\label{sec:moot}

\begin{table*}[ht]
\scriptsize
\renewcommand{\baselinestretch}{0.9}
\caption{Summary of datasets in the MOOT repository. ``x/y'' denotes the number of independent and dependent attributes.}
\label{datasets-summary}
\begin{tabular}{p{1cm}p{2.5cm}p{3.5cm}p{4.5cm}p{1.5cm}p{1cm}p{1cm}}
\# Datasets & Dataset Type                     & File Names                                                               & Primary Objective                                                     & x/y          & \# Rows       & Experts\\ \midrule
25          & \begin{tabular}[c]{@{}l@{}}Specific Software\\Configurations\end{tabular} & SS-A to SS-X, billing10k                                                 & Optimize software system settings                                     & 3-88/2-3   & 197–86,059 &   \\ 
12          & \begin{tabular}[c]{@{}l@{}}PromiseTune Software\\Configurations\end{tabular} & \begin{tabular}[c]{@{}l@{}} 7z, BDBC, HSQLDB, LLVM, PostgreSQL, \\ dconvert, deeparch, exastencils, javagc, \\ redis, storm, x264\end{tabular}                                                 & Software performance optimization                                     &  9-35/1  &  864-166,975 & \checkmark \\ \midrule
1           & Cloud                            & HSMGP num                                                                & Hazardous Software, Management Program data                           & 14/1         & 3,457     \\
1           & Cloud                            & Apache AllMeasurements                                                   & Apache server performance optimization                                & 9/1          & 192     \\
1           & Cloud                            & SQL AllMeasurements                                                      & SQL database tuning                                                   & 39/1         & 4,654    \\
1           & Cloud                            & X264 AllMeasurements                                                     & Video encoding optimization                                           & 16/1         & 1,153 \\
7           & Cloud                            & (rs—sol—wc)*                                                             & misc configuration tasks                                              & 3-6/1      & 196–3,840 \\ \midrule
35          & Software Project Health          & Health-ClosedIssues, -PRs, -Commits                                      & Predict project health and developer activity                         & 5/2-3      & 10,001 & \checkmark \\ \midrule
3           & Scrum                            & Scrum1k, Scrum10k, Scrum100k                                             & Configurations of the scrum feature model                             & 124/3      & 1,001–100,001 & \checkmark \\ \midrule
8           & Feature Models                   & FFM-*, FM-*                                                              & Optimize number of variables, constraints and Clause/Constraint ratio & 128-1,044/3 & 10,001 \\ \midrule
1 &	Software Process Model &	nasa93dem &	Optimize effort, defects, time and LOC	& 24/3 &	93 & \checkmark \\
1           & Software Process Model           & COC1000                                                                  & Optimize risk, effort, analyst experience, etc                        & 20/5         & 1,001 & \checkmark  \\
4           & Software Process Model           & POM3 (A–D)                                                               & Balancing idle rates, completion rates and cost                       & 9/3          & 501–20,001 & \checkmark\\
4           & Software Process Model    & XOMO (Flight, Ground, OSP)                                               & Optimizing risk, effort, defects, and time                            & 27/4         & 10,001 & \checkmark \\ \midrule
3           & Miscellaneous                             & auto93, Car\_price, Wine\_quality                                        & Miscellaneous                                                         & 5-38/2-5   & 205–1,600 & \checkmark\\ \midrule
4           & Behavioral                       & all\_players, student\_dropout,\newline HR-employeeAttrition, player\_statistics & Analyze and predict behavioral patterns                              & 26-55/1-3  & 82–17,738 & \checkmark \\ \midrule
4           & Financial                        & BankChurners, home\_data, Loan, \newline Telco-Churn                              & Financial analysis and prediction                                     & 19-77/2-5  & 1,460–20,000 \\ \midrule
3           & Human Health Data                & COVID19, Life\_Expectancy, \newline hospital\_Readmissions                        & Health-related analysis and prediction                                & 20-64/1-3  & 2,938–25,000 & \checkmark   \\ \midrule
2           & Reinforcement Learning           & A2C\_Acrobot, A2C\_CartPole                                              & Reinforcement learning tasks                                          & 9-11/3-4   & 224–318  & \checkmark     \\ \midrule
5           & Sales                            & accessories, dress-up, Marketing\_Analytics, socks, wallpaper            & Sales analysis and prediction                                         & 14-31/1-8  & 247–2,206 \\ \midrule
2	& Software testing	& test120, test600	& Optimize the class	& 9/1	& 5,161 & \checkmark\\ \midrule

127         & \textbf{Total}                            &                                                                          &                                                                       &              &        &      
\end{tabular}
\end{table*}

To evaluate our hypotheses, we utilize MOOT, a curated repository of software engineering optimization datasets. These datasets come from papers published in top SE venues such as 
the International Conference on Software Engineering \cite{chen2026promisetune, DBLP:conf/icse/WeberKSAS23,10172849,DBLP:conf/icse/HaZ19},
Foundations of SE (FSE) conference~\cite{nair2017using,DBLP:conf/sigsoft/JamshidiVKS18}
IEEE Trans. SE \cite{chen2025accuracy,xia2020sequential,krishna2020whence,Chen19,krall2015gale},
the Information Software Technology journal \cite{chen2018beyond,fu2016tuning},
Empirical Softw. Eng. \cite{hulse2025shaky, peng2023veer,guo2018data},
Mining Software Repositories \cite{nair18},
IEEE Access \cite{lustossa2024isneak},
 ACM Trans. SE Methodologies~\cite{lustosa2024learning} and the Automated Software Engineering Journal~\cite{nair2018faster}. The repository currently houses over 120 datasets spanning diverse domains, including real-world system traces and software process simulations. Table. \ref{datasets-summary} provides a summary of the datasets in MOOT.

\subsubsection{Dataset Selection and Stratification}
From this corpus, we curate a representative subset to form our experimental testbed. We enforce a selection criterion of \textbf{at least 10 datasets per complexity tier} to ensure generalizability across the dimensionality spectrum in accordance with prior work \cite{senthilkumar2024largelanguagemodelsimprove}:
\begin{itemize}
    \item \textbf{Low Dimensionality ($<6$ features):} Simple configuration tasks where LLMs historically perform well.
    \item \textbf{Medium Dimensionality ($6-11$ features):} Intermediate complexity tasks.
    \item \textbf{High Dimensionality ($>11$ features):} Complex, sparse landscapes where prior work indicates LLMs fail.
\end{itemize}

This selection strategy is designed to balance statistical power with economic constraints. Our study requires four distinct knowledge integration methods (plus baselines) executed 20 independent trials across datasets. Due to costs associated with high-volume API access for state-of-the-art LLMs, analyzing the entire repository is infeasible. By restricting our testbed to a stratified subset, we ensure sufficient statistical power to detect the "dimensional barrier" while managing API costs.

The final column of Table~\ref{datasets-summary} indicates the feasibility of recruitment for the H-DKP protocol. Datasets marked with a \checkmark indicate domains where Subject Matter Experts (SMEs) are accessible either locally at the NC State campus or through our direct academic network. This accessibility filter will serve as a secondary criterion when choosing datasets from MOOT for H-DKP.

\subsection{Participants: Subject Matter Experts (SMEs)}

We will recruit human experts following approval from the North Carolina State University Institutional Review Board (IRB). We classify this study as ``Minimal Risk'' regarding human subjects, as data collection is limited to professional email correspondence regarding technical domain knowledge, with no collection of sensitive personal data, health information, or identifiers linked to vulnerable populations.

\subsubsection{Recruitment Channels and Criteria}
We leverage the extensive industry connections of the NC State Computer Science Department, where many graduates have remained in the Research Triangle Park area for over a decade. Through the departmental alumni group, advisory board, and faculty contacts, we will conduct a targeted email campaign to identify qualified experts. Based on prior experience with similar SE surveys, we anticipate a response rate of 2-10\%. To achieve our target of at least one expert per dataset category, we plan to issue approximately 500 recruitment requests. \color{blue} 
We will conduct a pilot study with two experts local to the university to calibrate expert time commitment before wider recruitment.\color{black}

We define a qualified Subject Matter Expert (SME) based on meeting at least one of the following criteria:
\begin{itemize}
    \item \textbf{Primary Authorship:} Authors of the original research papers contributing datasets to the MOOT repository.
    \item \textbf{Project Maintenance:} Active maintainers or core contributors to the specific software systems under consideration.
    \textbf{Gold Standard Vetting:} For experts who are not primary authors or known maintainers, we will administer a brief "Gold Standard" questionnaire containing 
    7 multiple choice questions about the domain. These questions will focus on specific, non-obvious domain constraints found in the documentation. Only participants who correctly answer these control questions will be admitted to the study. \color{black}
\end{itemize}

\subsubsection{Contingency for Dropouts and Asynchronous Responses}
We recognize that expert availability is variable. Therefore, we adopt a flexible Asynchronous Iterative Protocol:
\begin{itemize}
    \item \textbf{Non-Consecutive Iterations:} The experimental design requires $T$ logical feedback iterations, not $T$ consecutive calendar days. Experts may respond at their own pace; the LLM state remains frozen until feedback is received.
    \item \textbf{Minimum Viable Threshold ($T_{min}$):} To ensure statistical validity, we establish a minimum threshold of $T_{min}=5$ iterations. Datasets where experts provide fewer than 5 feedback cycles will be excluded from the H-DKP specific analysis.
    \item \textbf{Variable $T$ Analysis:} We anticipate that the final number of iterations $T$ will vary across datasets (e.g., $5 \le T \le 10$). We pre-register an analysis to correlate the number of expert feedback rounds ($T$) with the final optimization improvement ($\Delta$ Chebyshev distance). This allows us to quantify the marginal utility of human effort (e.g., \textit{"Does performance plateau after $T=7$?"}).
\end{itemize}

\section{Analysis Plan \& Evaluation Criteria}

\subsection{Baseline Methods}
We will compare our methods against the following warm start baselines:

\begin{itemize}
    \item \textbf{Random Sampling (Random):} A naive baseline that selects samples uniformly at random from the search space. This establishes the lower bound of performance.
    \item \textbf{Gaussian Process Model (UCB\_GPM):} The current state-of-the-art for high-dimensional optimization in this domain. We use a Gaussian Process regressor with the Upper Confidence Bound (UCB) \cite{Srinivas_2012} acquisition function.
    \item \textbf{Standard LLM Warm Start (BS\_LLM):} Prior state-of-the-art few-shot prompting approach where the LLM is provided with 4 randomly selected examples (labeled as "Best" or "Rest") and basic feature metadata (name, type, median) \cite{senthilkumar2024largelanguagemodelsimprove}. This serves as our primary control to measure improvements from domain knowledge integration.
\end{itemize}

\subsection{Primary Metric: Chebyshev Distance}
We quantify the quality of generated warm starts using the Chebyshev distance to the optimal configuration as done in prior work \cite{senthilkumar2024largelanguagemodelsimprove}. 
Since objectives in software optimization often have vastly different scales, we first normalize all objective values $y_i$ to the range $[0, 1]$. The Chebyshev distance $D$ for a candidate solution $x$ is defined as:

\begin{equation}
D(x) = \max_{i \in \{1, \dots, m\}} (|f_i(x) - z_i^*|)
\end{equation}

Where $m$ is the number of objectives and $z_i^*$ is the ideal value (0 for minimization) for the $i$-th objective. A lower Chebyshev distance indicates a solution closer to the theoretical optimum.
We perform our experiments for 20 trials across each method (in H-DKP, the final prompt is used 20 times to generate warm starts). For each trial, we report the minimum Chebyshev distance achieved among the generated warm start examples. \color{black}

\subsection{Secondary Metrics}

\subsubsection{Generated Example Diversity}
To ensure the LLM is not simply generating identical variations of one good example, we measure the diversity of the generated set $\mathcal{E}_{gen}$. We calculate the \textit{Average Pairwise Euclidean Distance} between all generated vectors in the feature space. Higher diversity implies better exploration of the search space.

\subsubsection{Computational Cost (API Tokens)}
We track the economic feasibility of each method by logging the total number of input and output tokens consumed per trial. We report the average cost per successful warm start. This allows us to analyze the tradeoff between performance gains and the increased inference cost relative to the single-shot baseline.\color{black}

\subsection{Statistical Analysis: Scott-Knott and Effect Size}
To determine if our proposed methods provide a statistically significant improvement over the baseline, we employ the Scott-Knott Effect Size Difference (ESD) test.

\subsubsection{Scott-Knott Clustering Algorithm}
The Scott-Knott algorithm \cite{b399ed80-3ccc-30cf-af0d-87bbdad7ade6} recursively partitions the set of treatment means into two subsets to maximize the difference between groups. The splitting criterion maximizes the \textit{Between-Group Sum of Squares} ($B_0$).
For a set of treatments with sizes $N_1$ and $N_2$ and sums of responses $T_1$ and $T_2$, the algorithm seeks a partition that maximizes:

\begin{equation}
B_0 = \frac{T_1^2}{N_1} + \frac{T_2^2}{N_2} - \frac{(T_1 + T_2)^2}{N_1 + N_2}
\end{equation}

The algorithm follows these steps:
\begin{enumerate}
    \item \textbf{Sort:} Order the treatment distributions by their median Chebyshev distance.
    \item \textbf{Split:} Identify the partition point that yields the maximum $B_0$.
    \item \textbf{Significance Test:} Check if the split is statistically significant using a bootstrap sampling method (to avoid assumptions of normality).
    \item \textbf{Effect Size Check (ESD):} Even if significant, the split is rejected if the magnitude of the difference is negligible (Cliff's Delta $< 0.147$). This ensures distinct ranks represent practically meaningful differences.
    \item \textbf{Recurse:} If the split is valid, recursively apply the procedure to each subgroup; otherwise, terminate and group the treatments into a single rank.
\end{enumerate}

\subsubsection{Effect Size (Cliff's Delta)}
To ensure that observed differences are not just statistically significant but practically meaningful, we calculate Cliff's Delta \cite{Cliff1993DominanceSO} ($\delta$), a non-parametric effect size measure. We interpret the magnitude of difference between the proposed method and the baseline as follows:
\begin{itemize}
    \item $|\delta| < 0.147$: Negligible
    \item $0.147 \leq |\delta| < 0.33$: Small
    \item $0.33 \leq |\delta| < 0.474$: Medium
    \item $|\delta| \geq 0.474$: Large
\end{itemize}
We consider a hypothesis validated only if the proposed method achieves a better Scott-Knott rank \textit{and} shows at least a "Small" effect size improvement over the BS\_LLM baseline.

\section{Methods \& Execution Plan}

\subsection{Human-in-the-Loop Domain Knowledge Prompting (H-DKP)}

\subsubsection{Overview and Rationale}
While LLMs can parse documentation, they lack the tacit knowledge possessed by domain experts (unwritten rules, edge cases, and intuition gained through experience). We propose \textbf{Human-in-the-Loop Domain Knowledge Prompting (H-DKP)}, a methodology to extract this tacit knowledge through a structured, asynchronous dialogue with human experts.
Unlike static few-shot prompting, H-DKP treats the prompt construction as an iterative software design process, evolving the LLM's "mental model" of the domain over a fixed time window. The process is outlined in Algorithm \ref{alg:hdkp}.

\subsubsection{Expert Recruitment and Elicitation Protocol}
To execute H-DKP, we will identify and recruit domain experts for the datasets in the MOOT repository as outlined previously. To minimize expert cognitive load, we utilize the Recognition over Recall principle. Experts will not be asked to write prompts. Instead, they will critique the LLM's outputs.

\subsubsection{The 10-Day Iterative Refinement Sprint}
We employ an \textbf{Asynchronous Iterative Refinement (AIR)} protocol. For a period of 10 consecutive days, each expert engages in a daily feedback loop with the LLM.

\begin{itemize}
    \item \textbf{Day 1 (Initialization):} The LLM generates an initial ``Belief State'' (hypothesized constraints and feature relationships) 
    based on documentation about the data, variable names and problem objectives.\color{black} This is sent to the expert for a baseline validity check (Valid/Invalid/Modify).
    \item \textbf{Days 2-9 (The Feedback Loop):} 
    Each day, we run the current Prompt State to generate warm start configurations. We identify the ``Most Confusing Failure''—a configuration the LLM predicted would be optimal but which performed poorly in reality.
    The expert receives a structured email containing:
    \begin{enumerate}
        \item The current rule set the LLM is following.
        \item The specific failure case (e.g., \textit{"The model set `threads=100` expecting high throughput, but latency spiked."}).
        \item A single question: \textit{"What domain rule is the model missing that explains this failure?"}
    \end{enumerate}
    The expert's email reply is parsed and appended to the prompt context for the next day's run.
    \item \textbf{Day 10 (Finalization):} The accumulated knowledge base is frozen and used for the final experimental evaluation.
\end{itemize}

\begin{algorithm}
\caption{H-DKP via Asynchronous Expert Feedback}
\label{alg:hdkp}
\begin{algorithmic}[1]
\REQUIRE Dataset $\mathcal{D}$, Human Expert $\mathcal{H}$
\REQUIRE Duration $T=10$ days
\ENSURE Optimized Knowledge Base $K_{final}$ and Warm Starts $\mathcal{E}$

\STATE \textbf{Day 1: Bootstrapping}
\STATE $K_1 \gets \text{LLM}(\text{Docs}, \text{"Propose constraints"}) $
\STATE Send Email($\mathcal{H}$, "Verify these baseline constraints: $K_1$")
\STATE $K_1 \gets \text{Update}(K_1, \text{EmailReply}(\mathcal{H}))$

\FOR{$t = 2$ to $T$}
    \STATE \COMMENT{Generate candidates using current knowledge}
    \STATE $\mathcal{E}_t \gets \text{LLM}(\text{prompt}=K_{t-1})$
    \STATE Evaluate $\mathcal{E}_t$ against ground truth $\mathcal{D}$
    
    \STATE \COMMENT{Identify knowledge gap}
    \STATE $e_{fail} \gets \text{FindMaxError}(\mathcal{E}_t)$ \COMMENT{High confidence, low reward}
    
    \STATE \COMMENT{Daily Asynchronous Query}
    \STATE $\text{Query} \gets \text{"Model believed } e_{fail} \text{ was optimal. Why did it fail?"}$
    \STATE Send Email($\mathcal{H}$, $\text{Query}$)
    
    \STATE \COMMENT{Wait for asynchronous reply}
    \STATE $\text{Feedback}_t \gets \text{EmailReply}(\mathcal{H})$
    \STATE $K_t \gets K_{t-1} \cup \text{Feedback}_t$
\ENDFOR

\STATE \textbf{Final Evaluation}
\STATE $\mathcal{E}_{final} \gets \text{LLM}(\text{prompt}=K_{10})$
\STATE \textbf{return} $\mathcal{E}_{final}$
\end{algorithmic}
\end{algorithm}

\subsubsection{H-DKP analysis consideration}
Given the nature of human studies and mercurial nature of expert recruitment, we do not anticipate our analysis with H-DKP to span across all datasets used in the other discussed methods. We will restrict our comparison of H-DKP and other algorithms only for datasets where the study is completed with an expert without dropout. \color{black}

\subsection{Adaptive Multi-Stage Prompting (AMP)}

\subsubsection{Overview and Rationale}
Standard "single-shot" prompting treats optimization as a pattern-matching task, asking the LLM to generate solutions immediately after seeing a few examples.
We propose \textbf{Adaptive Multi-Stage Prompting (AMP)}, a sequential reasoning pipeline that forces the LLM to explicitly articulate its "mental model" of the optimization landscape before generating configuration values. By separating analysis from generation, AMP aims to reduce the logical inconsistencies and constraint violations common in single-shot warm starts.

\subsubsection{Stage 1: Analysis}
The first stage functions as a filter to distinguish signal from noise. The LLM is provided with dataset metadata (feature names, types, ranges) and the initial few-shot examples. It is prompted to output a structured analysis identifying:
\begin{enumerate}
    \item \textbf{Feature Ranking:} A prioritized list of the 3-5 most influential features driving the objective values.
    \item \textbf{Tradeoff Identification:} Explicit notes on conflicting objectives observed in the few-shot examples.
    \item \textbf{Directionality:} The hypothesized direction of improvement for continuous variables.
\end{enumerate}

\subsubsection{Stage 2: Constraint Discovery}
Using the analysis from Stage 1, the LLM infers explicit boundaries for valid configurations. This stage distinguishes between:
\begin{itemize}
    \item \textbf{Hard Constraints:} Inviolable rules based on physical or logical limits (e.g., \textit{"parallel\_threads cannot exceed available cpu\_cores"}).
    \item \textbf{Soft Constraints:} Heuristic preferences that generally lead to better outcomes but may be violated for exploration.
\end{itemize}
We expect that explicitly generating these rules will reduce the search space for the subsequent generation step and provide a "rulebook" for the validation stage.

\subsubsection{Stage 3: Constrained Generation}
The third stage performs the actual warm start generation. Unlike the baseline single-shot approach, the prompt for this stage is dynamically constructed to include the prioritized Feature List (from Stage 1) and the Validation Rules (from Stage 2) as strict instructions.
The LLM is tasked to generate configurations that optimize the identified key features while strictly adhering to the discovered hard constraints.

\subsubsection{Stage 4: Self-Validation}
In the final stage, the LLM acts as a critic. It reviews its own generated configurations against the constraint set defined in Stage 2.
\begin{itemize}
    \item \textbf{Verification:} Each generated example is checked for 
    strict logical consistency based on the generated constraints.\color{black}
    \item \textbf{Refinement:} If a configuration violates a hard constraint, the model is prompted to revise the specific value while preserving the rest of the configuration.
\end{itemize}

\subsubsection{AMP Ablation study}
To study the effect of the different strategies for prompting the LLM to decipher knowledge we will study the various stages in 3 experimental conditions: Condition 1 (AMP-2): Analysis + Generation, Condition 2 (AMP-3): Analysis + Constraints + Generation \& Condition 3 (AMP-4): Full 4-stage pipeline.

\subsection{Dimension-Aware Progressive Refinement (DAPR)}
\subsubsection{Feature Importance Ranking}
To minimize possible bias and failures of specific methods, we propose feature importance calculation through 3 different statistical methods - Spearman coefficient, mutual information \& feature importance calculated through random forest. The calculated values through these methods will then be normalized and averaged to calculate the final feature importance score. It is to be noted that this calculation will only be performed with the few-shot samples random chosen as knowledge for the LLM.

\subsubsection{Progressive Expansion Algorithm}

The progressive refinement portion of DAPR, as illustrated in Algorithm. \ref{alg:dapr}, begins by initializing the current feature set $\chi_{cur}$with the top
$k$ most important features (line 2). At each iteration, the algorithm projects four random examples onto the current reduced subspace (line 4) and prompts the LLM to generate optimized configurations using only these features (line 5). Generated examples are mapped to their nearest neighbors 
(following the methodology in prior work \cite{pmlr-v188-pfisterer22a, zela2022surrogatenasbenchmarksgoing}) \color{black} in the full dataset to obtain labels (lines 6-8), and the best configuration is tracked across iterations (lines 9-12). The feature space is then progressively expanded by adding the next $s$ most important features (lines 13-14), with newly added features anchored to values from the current best configuration to maintain continuity (lines 15-17). This process repeats until all $n$ features are included. Finally, the algorithm generates warm start examples in the full dimensional space, using the best configuration found during progressive refinement as an anchor point (lines 18-20).

\begin{algorithm}
\caption{Dimension-Aware Progressive Refinement (DAPR)}
\begin{algorithmic}[1]
\REQUIRE Dataset $\mathcal{D}$ with features $\mathcal{X} = \{x_1, \ldots, x_n\}$
\REQUIRE Ranked features $\mathcal{F} = [f_1, \ldots, f_n]$, initial size $k$, step $s$
\ENSURE Warm start examples in full $n$-dimensional space
\STATE $\mathcal{X}_{\text{cur}} \gets \{f_1, \ldots, f_k\}$; $\text{best} \gets \text{null}$
\WHILE{$|\mathcal{X}_{\text{cur}}| < n$}
    \STATE $\mathcal{E}_{\text{fs}} \gets$ project 4 random samples onto $\mathcal{X}_{\text{cur}}$
    \STATE $\mathcal{E}_{\text{gen}} \gets \text{LLM}(\mathcal{E}_{\text{fs}}, \texttt{"Optimize on } \mathcal{X}_{\text{cur}}\texttt{"})$
    \FOR{$e \in \mathcal{E}_{\text{gen}}$}
        \STATE $e_{\text{full}} \gets$ nearest neighbor of $e$ in $\mathcal{D}$
        \STATE Evaluate Chebyshev$(e_{\text{full}})$
    \ENDFOR
    \STATE $e^* \gets \arg\min_{e \in \mathcal{E}_{\text{gen}}} \text{Chebyshev}(e_{\text{full}})$
    \IF{$\text{best} = \text{null}$ or $\text{Chebyshev}(e^*) < \text{Chebyshev}(\text{best})$}
        \STATE $\text{best} \gets e^*$
    \ENDIF
    \STATE $d \gets |\mathcal{X}_{\text{cur}}|$;
           $\mathcal{X}_{\text{new}} \gets \{f_{d+1}, \ldots, f_{\min(d+s,n)}\}$
    \STATE $\mathcal{X}_{\text{cur}} \gets \mathcal{X}_{\text{cur}} \cup \mathcal{X}_{\text{new}}$
    \FOR{$x_i \in \mathcal{X}_{\text{new}}$}
        \STATE Anchor $x_i$ to value from $\text{best}$ (or $\text{median}(x_i)$ if unavailable)
    \ENDFOR
\ENDWHILE
\STATE $\mathcal{E}_{\text{fs}} \gets$ 4 random samples in full space
\STATE $\mathcal{E}_{\text{final}} \gets \text{LLM}(\mathcal{E}_{\text{fs}}, \text{anchored to best})$
\STATE \textbf{return} $\mathcal{E}_{\text{final}}$
\end{algorithmic}
\label{alg:dapr}
\end{algorithm}

\subsection{Hybrid Knowledge-Model Approach (HKMA)}

\subsubsection{Overview and Rationale} We posit that LLMs and statistical models suffer from orthogonal blind spots: LLMs possess semantic understanding but hallucinate quantitative relationships, while Bayesian models (like TPE) identify quantitative patterns but lack semantic causal reasoning. HKMA bridges this gap using a lightweight statistical model to "scout" the terrain and identify empirical priors, which are then fed to the LLM to ground its generation in observed reality.

\subsubsection{Phase 1: Statistical Scouting} 
Before invoking the LLM, we perform a rapid, low-budget exploration using the Tree-structured Parzen Estimator (TPE). We allocate a small "scouting budget" ($B_{scout} = 10$) to perform a rapid exploration of the search space using TPE. We specifically utilize the TPE exploit capability to quickly identify high-performing regions versus low-performing ones. By comparing the distribution of the top-performing configurations ($S_{best}$) against the remaining samples ($S_{rest}$), we extract Empirical Priors. These priors are formalized as natural language descriptions of observed phenomena, such as directional trends and boundary conditions.

\subsubsection{Phase 2: Retrieval-Augmented Synthesis}

We employ Retrieval-Augmented Generation (RAG) to provide semantic context for the observed statistical patterns.
We index the domain documentation and academic literature collected from the MOOT repository into a vector store.\color{black} Using the extracted Empirical Priors as search queries, we retrieve relevant documentation that explains the physical or logical mechanisms behind the statistics (e.g., querying "Why does high buffer size improve throughput?" to retrieve specific memory management docs). The final prompt to the LLM constructs a synthesis task: the model is provided with the Empirical Evidence and the Semantic Explanation, and is tasked with generating warm start configurations that satisfy both.

\subsubsection{Study design and ablation}
We plan to ablate the RAG and scouting phases by using them separately and together to understand the implication of these treatments on the quality of warm starts generated.

\printbibliography

\newcommand{\reviewercomment}[1]{
    \begin{framed}
    \noindent \textit{\textbf{Comment:} #1}
    \end{framed}
}

\newcommand{\response}[1]{
    \noindent \textbf{Response:} #1 \vspace{1em}
}

\newcommand{\change}[1]{
    \noindent \textcolor{blue}{\textbf{Action Taken/Revision:}} #1 \vspace{1.5em}
}

\end{document}